\documentclass{myws-p10x7}
\usepackage{amssymb}
\begin{document}

\title{Bose-Einstein Correlations in W-pair events at LEP}
\author{A. Valassi}
\address{
CERN, EP Division, 1211 Gen\`eve 23, Switzerland\\
E-mail: Andrea.Valassi@cern.ch
}

\twocolumn[
\maketitle\abstract{
An overview 
of measurements of Bose-Einstein correlations 
in W-pair events at LEP is given.
The results presented 
are based on data collected 
at centre-of-mass energies between 172 and 202 GeV.
The review concentrates on 
the search for Bose-Einstein correlations  
between pions from different W's
in $\mathrm{e^+e^-\rightarrow W^+W^-\rightarrow q\bar{q}q\bar{q}}$ events.
No agreement is reached in the results of the four experiments.
}]

\section{Introduction}
\vspace*{-1.5mm}
\vspace*{-0.8mm}
\vspace*{-0.3mm}
The existence of 
Bose-Einstein correlations (BEC)
in reactions leading to hadronic final states 
is well known.
BEC are observed as an enhancement of the production 
of multiple identical bosons close in momentum space,
first reported for pairs of charged pions 
in $\mathrm{p\bar{p}}$ collisions\cite{gold}.
They have also been studied for two or more identical bosons
in hadronic Z decays at LEP\cite{lepuno},
including $\pi^\pm$, K$^0_\mathrm{s}$ and K$^\pm$.

This talk reviews recent studies
of BEC\cite{aleph,delphi,l3,opal}
for charged pion pairs produced 
in $\mathrm{e^+e^-\rightarrow W^+W^-}$ events at LEP.
``Intra-W'' BEC (or BEI, BEC Inside a W)
between pions from the same $\mathrm{W\rightarrow q\bar{q}}$ decay
are observed unambiguously.
As expected,
these closely resemble BEC 
in $\mathrm{Z\rightarrow q\bar{q}}$ decays, 
when only udsc quark flavours are considered.

Since the average separation between the two W decays at LEP2 
is $\lesssim0.1$~fm,
smaller than the hadronisation scale of $\sim1$~fm,
``inter-W'' BEC (or BEB, BEC Between W's)
could also be expected between pions 
from different W's 
in $\mathrm{WW\rightarrow q\bar{q}q\bar{q}}$ (4q) events.
The theoretical framework is, however, still unclear.
In the absence of exact nonperturbative QCD calculations
for the symmetrized production amplitude 
of multiple mesons from hadronic W decays, BEC are 
described only by phenomenological models.
Many models exist, with contradictory predictions\cite{lonnblad}
about BEB.
This note will therefore concentrate
on the experimental measurements of BEB in W-pair events at LEP.

The question whether BEB exist is particularly important as the 
cross-talk between the two W decays 
could bias the W mass ($m_W$) measurement in 4q events.
Different models of BEC predict
different values for the shift\cite{lonnblad}.
Using MC events with full detector simulation,
the LEP experiments obtain\cite{fourwmasses} 
$m_W$ biases 
from 20 to 67 MeV,
conservatively taken as systematic errors 
on the individual measurements.
In the LEP combination,
a common BEC systematic error of 25 MeV in the 4q channel 
is assumed\cite{lepwmass}.
The impact of this error 
is more than halved when combining 
4q events to $\mathrm{WW\rightarrow\ell\nu{q}\bar{q}}$ (2q) events.
While presently the BEC error is not the limiting factor to the 
LEP measurement of $m_W$,
a better understanding of BEC in W decays is important in view of the 
future reduction of the other, currently larger, 
systematics.

\newcommand{\be}{\begin{equation}}
\newcommand{\ee}{\end{equation}}
\vspace*{-3mm}
\section{Overview of analysis methods}
\vspace*{-2mm}
BEC in identical boson pairs
are often studied 
by the two-particle correlation function
\be
R(Q) = \frac{\rho(Q)}{\rho_0(Q)} \quad.
\label{eq:rq}
\ee
Here,
$\rho$ is the two-boson density in the presence of BEC,
while the reference $\rho_0$
should ideally describe pair-production if there were no BEC.
Since BEC are largest at small four-momentum difference
$Q=\sqrt{-(p_1-p_2)^2}$,
in a simplified approach 
the effect can be studied in terms 
of this variable alone\cite{bowler}.

In the studies reviewed 
in this note,
Bose-Einstein (BE) enhancements of production rates are parametrised 
using functions like
\be
R(Q) \sim (1 + \lambda e^{-r^2 Q^2}) \quad,
\label{eq:fit}
\ee
which describe
source distributions of radius $r$
and BE ``strength'' $\lambda$.
Fit results for $r$ and $\lambda$ 
will not be compared here, since 
the various analyses differ in 
detector acceptance, choice of fit function and definition of $R$.

The choice of the reference 
$\rho_0$ and the definition of $R$ are distinctive features 
of each BEC analysis.
The simplest 
is to derive $\rho_0$ from
MC events with no BEC (``standard MC'').
Data can also be used: for instance, 
the density
for unlike-sign pion pairs, $\rho^{+-}$,
is often taken as a reference to study BEC 
in like-sign pairs, $\rho^{++,--}$.
Since unlike-sign pion pairs are not free from 
correlations other than BEC,
such as those due to resonance decays,
the correlation function is frequently computed as a double ratio of the form
\be
\vspace*{-0.6mm}
R^\mathrm{d.r.} (Q) = \frac 
           {\left(\rho^{++,--}/\rho^{+-}\right)^\mathrm{\;data}
            \hspace*{1.05cm}}
           {\left(\rho^{++,--}/\rho^{+-}
            \right)^\mathrm{\;standard\;MC}} ,
\label{eq:double}
\vspace*{-0.mm}
\ee
where resonance correlations in the MC cancel those in data.
Additional corrections 
may be introduced to 
describe
final-state Coulomb interactions\cite{gyulassi},
different in like- and unlike-sign pairs and
usually absent in the MC.

In the study of BEB, 
instead of comparing 4q data to a reference with no BEC at all,
one can finally define\cite{kittel} a 
function like
\be
\vspace*{-0.6mm}
R^\mathrm{mix} (Q) = \frac 
   {(\rho^{++,--})^\mathrm{\;4q}\hspace*{1.1cm}}
   {(\rho^{++,--})^\mathrm{\;``mixed\mbox{\scriptsize''}\;2q}} \quad,
\label{eq:mixed}
\vspace*{-0.mm}
\ee
where a reference with BEI and no BEB is obtained by 
``mixing'' hadronically decaying W's from pairs 
of different 2q data events.
Different mixing procedures exist;
in general, 
the two W's, chosen of opposite charges,
are boosted to be approximately 
back-to-back in the lab frame,
as in 4q events.
A double-ratio version of Eq.~(\ref{eq:mixed}) can also be used.

\vspace*{-2.1mm}
\section{Experimental results at LEP2}
\vspace*{-2.7mm}
The analyses presented 
use data collected by the four 
experiments 
at 172--202 GeV
(Tab.~\ref{tab:data}).
Signal efficiency and background contamination
are 70--90\% and 20\% (mainly $\mathrm{q\bar{q}}$) for 4q,
50--75\% and 6\% for 2q events.
Up to $\sim$7000 selected WW events per experiment 
are used, including both channels.

\begin{table}[t]
\vspace*{-1mm}
\begin{center}
\begin{tabular}{|l|c|c|c|} 
\hline 
Exper. & $\sqrt{s}$ [GeV] & $\cal{L}$ [pb$^{-1}$] & Status \\
\hline
ALEPH\protect\cite{aleph}   & 172--202 & 479 & Prel. \\
DELPHI\protect\cite{delphi} & 183--202 & 437 & Prel. \\
L3\protect\cite{l3}         & 189      & 177 & Publ. \\
OPAL\protect\cite{opal}     & 172--189 & 250 & Prel. \\
\hline
\end{tabular}
\caption{Data samples used in the analyses presented.
         The L3 analysis is submitted for publication,
         those from the other experiments 
         are preliminary.
         }
\label{tab:data}
\vspace*{-7mm}
\end{center}
\end{table}

All analyses are performed on 
pairs of like-sign charged pions,
with generally loose pion selection criteria.
In each event, every pion enters several pairs,
which introduces bin-to-bin correlations 
in some distributions.
Where necessary,
these are corrected for
by all experiments, 
using various techniques.

The four experiments simulate BEC 
using different MC models,
differently tuned.
WW MC events 
with BEC are needed for model-dependent analyses
or cross-checks of model-independent assumptions.
BEC also exist in the q$\mathrm{\bar{q}}$ background,
although they may differ in the four experiments
depending on the anti-b tagging criteria used in the selection.
The simulation of BEC in $\mathrm{q\bar{q}}$ events selected as 4q
is especially important, as\cite{aleph} they
look more similar to BEC 
for 4q events 
in BEI+BEB MC than in BEI-only MC.

\begin{figure}[h]
 \vspace*{-0.8cm}
 \begin{center}
 \mbox{\hspace*{0.3cm}\epsfig{file=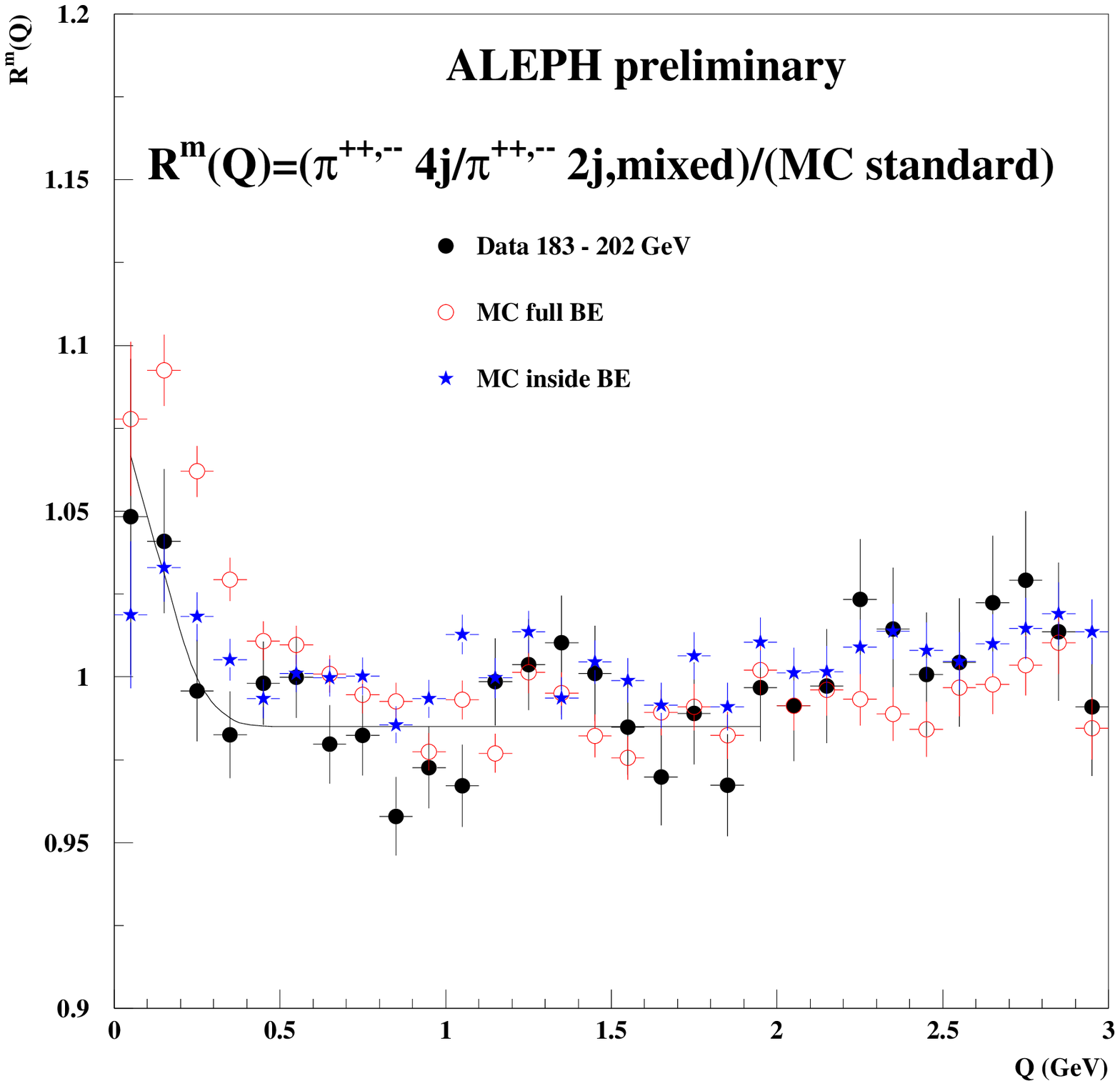,width=6.2cm}}
 \end{center}
 \vspace*{-0.4cm}
 \caption{ALEPH\protect\cite{aleph} distributions of $R^\mathrm{m}$
          for data,
          BEI-only MC (``inside'') and BEI+BEB MC (``full'').}
 \label{fig:aleph}
 \vspace*{-0.2cm}
\end{figure}

In ALEPH\cite{aleph},
a double ratio $R^*$ of like- and unlike-sign pion pairs for data 
over
standard MC is used,
similar to that of Eq.~(\ref{eq:double}).
The analysis of BEB is based
on the comparison of $R^*$ for data to that expected from MC
with BEI-only or BEI+BEB, 
simulated according to a specific model of BEC tuned on Z data.
Background is added to
the WW signal
in computing $R^*$ for the MC.
Fits to the $R^*(Q)$ distributions are performed
using a function similar to Eq.~(\ref{eq:fit}).
The results are compared in terms
of the integral of the BE 
signal,
$I=\int_0^\infty\lambda{e^{-r^2Q^2}dQ}=\frac{\sqrt{\pi}}{2}\frac{\lambda}{r}$.
The value for $I$ in the BEI-only MC is compatible with that in data, while 
$I$ in data and in the MC with BEI+BEB are inconsistent
at the level of 2.2$\sigma$.
This includes systematics,
dominated by the tuning of BEC MC parameters using Z data.
It is concluded that data support the BEI model considered,
while the 
BEB model is disfavoured.
An alternative analysis, 
using a double ratio $R^\mathrm{m}$ of 4q and mixed 2q events
in data over standard MC,
yields qualitatively similar results
(Fig.~\ref{fig:aleph}).

\begin{figure}[t]
 \begin{center}
 \vspace*{-1.2cm}
 \mbox{\hspace*{0.4cm}\epsfig{file=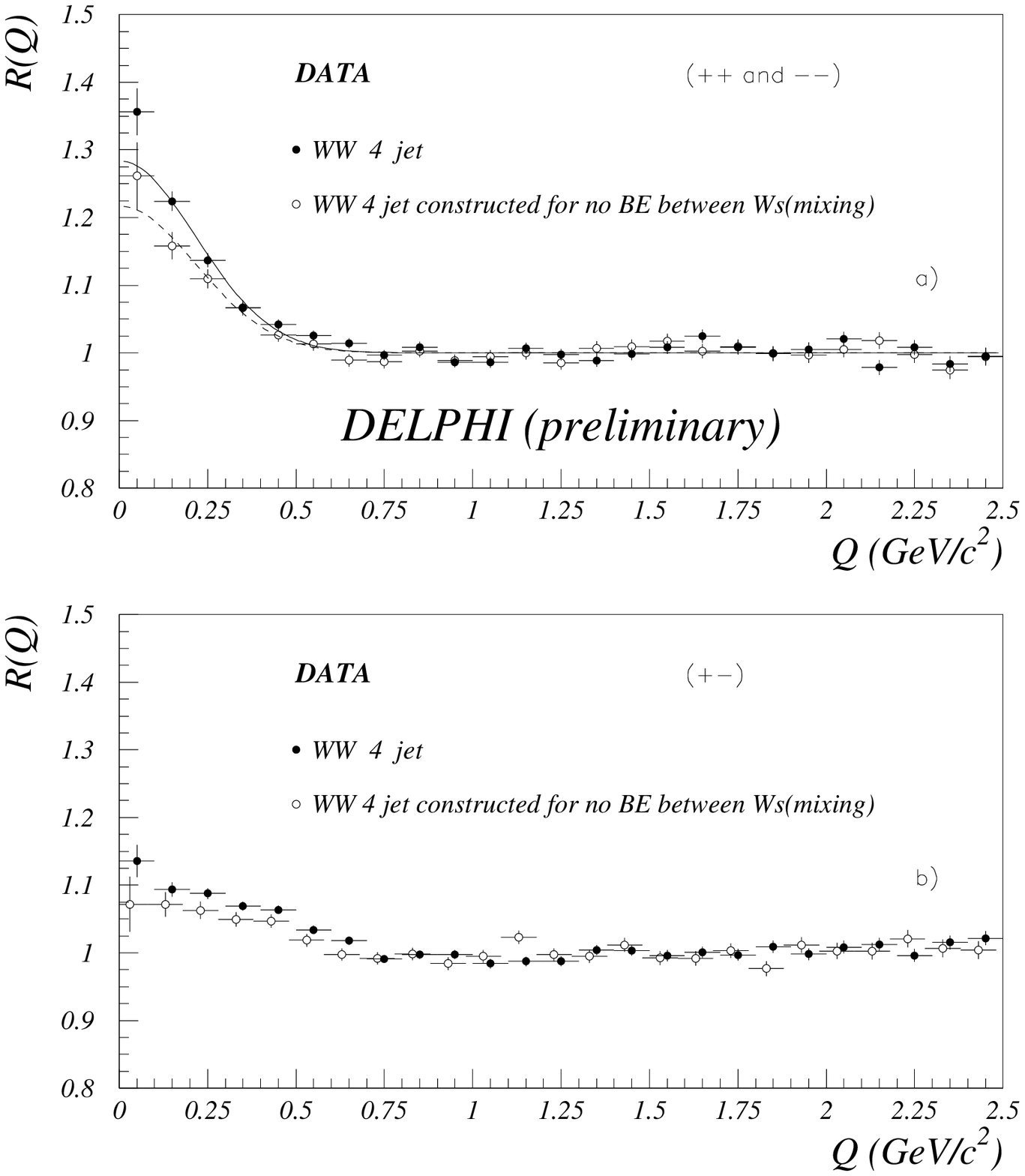,width=6.8cm}}
 \end{center}
 \vspace*{-0.4cm}
 \caption{DELPHI\protect\cite{delphi} distributions of 
          $R_\mathrm{4q}$ (black circles) 
          and $R_\mathrm{4q}^\mathrm{mix}$ (hollow circles)
          for a) like-sign and b) unlike-sign pion pairs,
          as a function of $Q$.}
 \label{fig:delphi}
 \vspace*{-0.4cm}
\end{figure}

In the DELPHI analysis\cite{delphi},
also presented 
in another talk\cite{puk},
correlation functions 
for like-sign pion pairs
$R_\mathrm{4q}$ and $R_\mathrm{4q}^\mathrm{mix}$ are built for 
both 4q events and mixed 2q events,
as single ratios over standard MC
(Fig.~\ref{fig:delphi}a).
Unlike-sign pairs are also used
for qualitative comparisons (Fig.~\ref{fig:delphi}b).
Expected backgrounds are 
subtracted from data
in computing
$R_\mathrm{4q}$ and $R_\mathrm{4q}^\mathrm{mix}$.
By construction,
$R_\mathrm{4q}(Q)$ and $R_\mathrm{4q}^\mathrm{mix}(Q)$ are
thus expected to be equal (different)
in the absence (presence) of BEB.
Signal MC events with 
BEC (BEI-only, or BEI+BEB)
are used to verify these hypotheses and for systematic studies.
A simultaneous fit to $R_\mathrm{4q}(Q)$ and $R_\mathrm{4q}^\mathrm{mix}(Q)$
determines a common BE source radius and two different BE strengths,
which are expected to be equal in the absence of BEB.
The fit to their difference 
$\Delta\lambda^\mathrm{mix} =
 \lambda_\mathrm{4q}-\lambda_\mathrm{4q}^\mathrm{mix}$
yields
\be
\Delta\lambda^\mathrm{mix} = 0.062 \pm 0.025 \pm 0.021 \quad.
\ee
It is concluded that
data support BEB at the level of $\sim2\sigma$.
The largest systematic errors 
come from the check that $\Delta\lambda^\mathrm{mix}=0$ in BEI-only 
MC, and from bin-to-bin correlations.

\begin{figure}[t]
 \begin{center}
 \vspace*{-2mm}
 \vspace*{-4.3cm}
 \mbox{\hspace*{-0.4cm}\epsfig{file=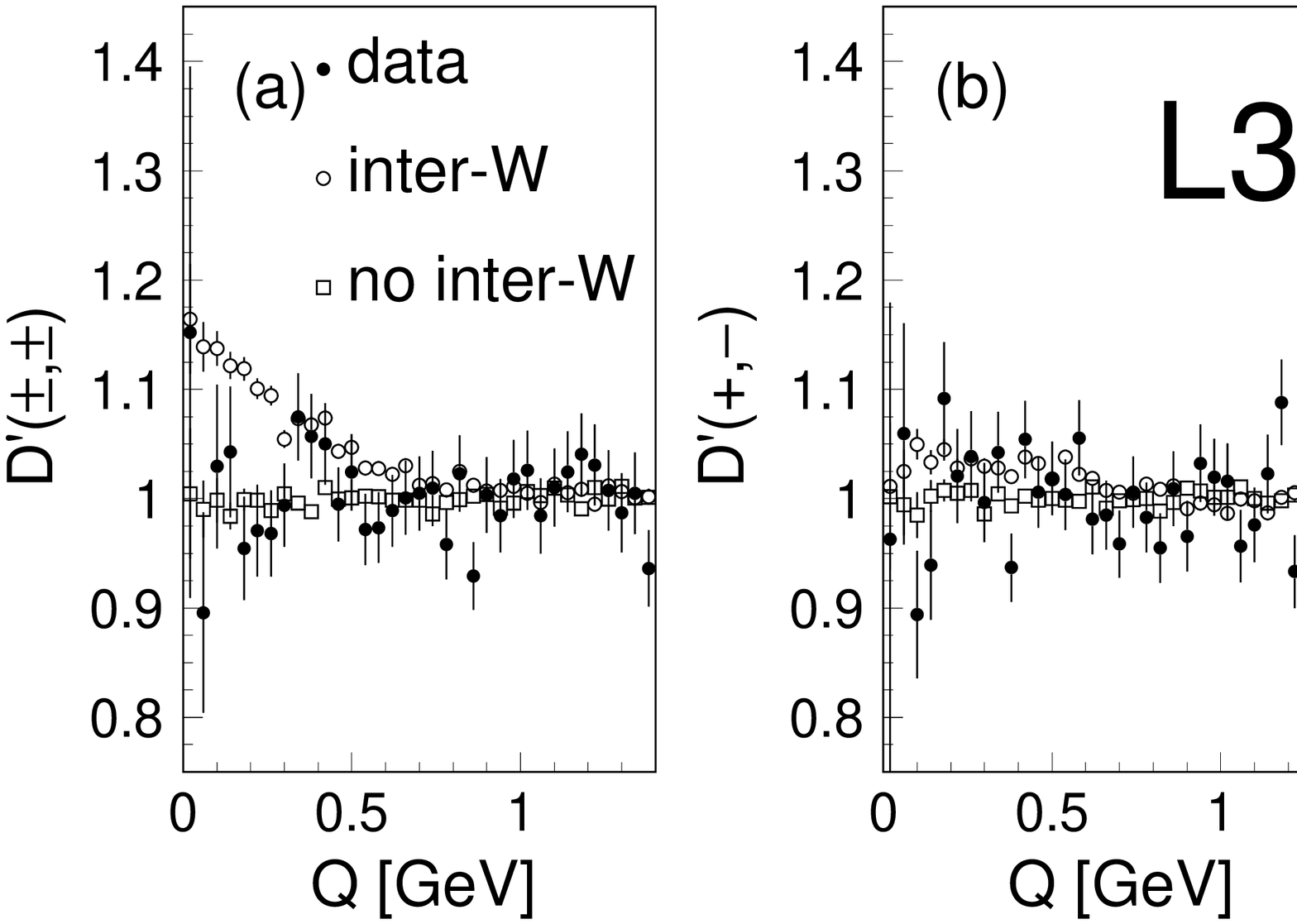,width=7.8cm}}
 \end{center}
 \vspace*{-0.8cm}
 \caption{L3\protect\cite{l3} distributions of $D'$ 
          for like-sign (a) and unlike-sign (b) pion pairs.
          Data are compared 
          to BEI-only (``intra-W'') and BEI+BEB (``inter-W'') MC.}
 \label{fig:l3}
 \vspace*{-0.4cm}
\end{figure}

In L3\cite{l3},
the correlation functions 
for like-sign charged pions are built taking 
mixed 2q events as a reference.
A single ratio $D$ is built 
as in Eq.~(\ref{eq:mixed}).
The main analysis uses a double ratio $D'$, where 
$D$ for data is divided by that for BEI-only MC events.
Backgrounds are subtracted from data
in the calculation of $D'$.
By construction, 
it is thus expected that $D'(Q)=1$
in the absence of BEB.
Signal MC with BEI-only is used to verify this hypothesis 
(Fig.~\ref{fig:l3}a).
Unlike-sign pairs are also used for qualitative comparison
(Fig.~\ref{fig:l3}b).
The $D'(Q)$ distribution is fitted
for the BEB strength $\Lambda$, expecting $\Lambda=0$ in the absence of BEB.
The fit yields
\be
\Lambda = 0.001 \pm 0.026 \pm 0.015 \quad,
\ee
indicating
that data are compatible with the absence of BEB.
Systematics are dominated by 
the selection of charged tracks
and the inclusion of the low-purity $\tau\nu\mathrm{q\bar{q}}$ channel.
The fit is repeated for MC events with a model of BEB,
yielding $\Lambda = 0.127\pm0.007$, where only statistical errors are given.
The BEB model considered is disfavoured by more than 4$\sigma$.

\begin{figure}[b]
 \vspace*{-0.9cm}
 \begin{center}
 \mbox{\hspace*{-0.35cm}\vspace*{0.0cm}
       \epsfig{file=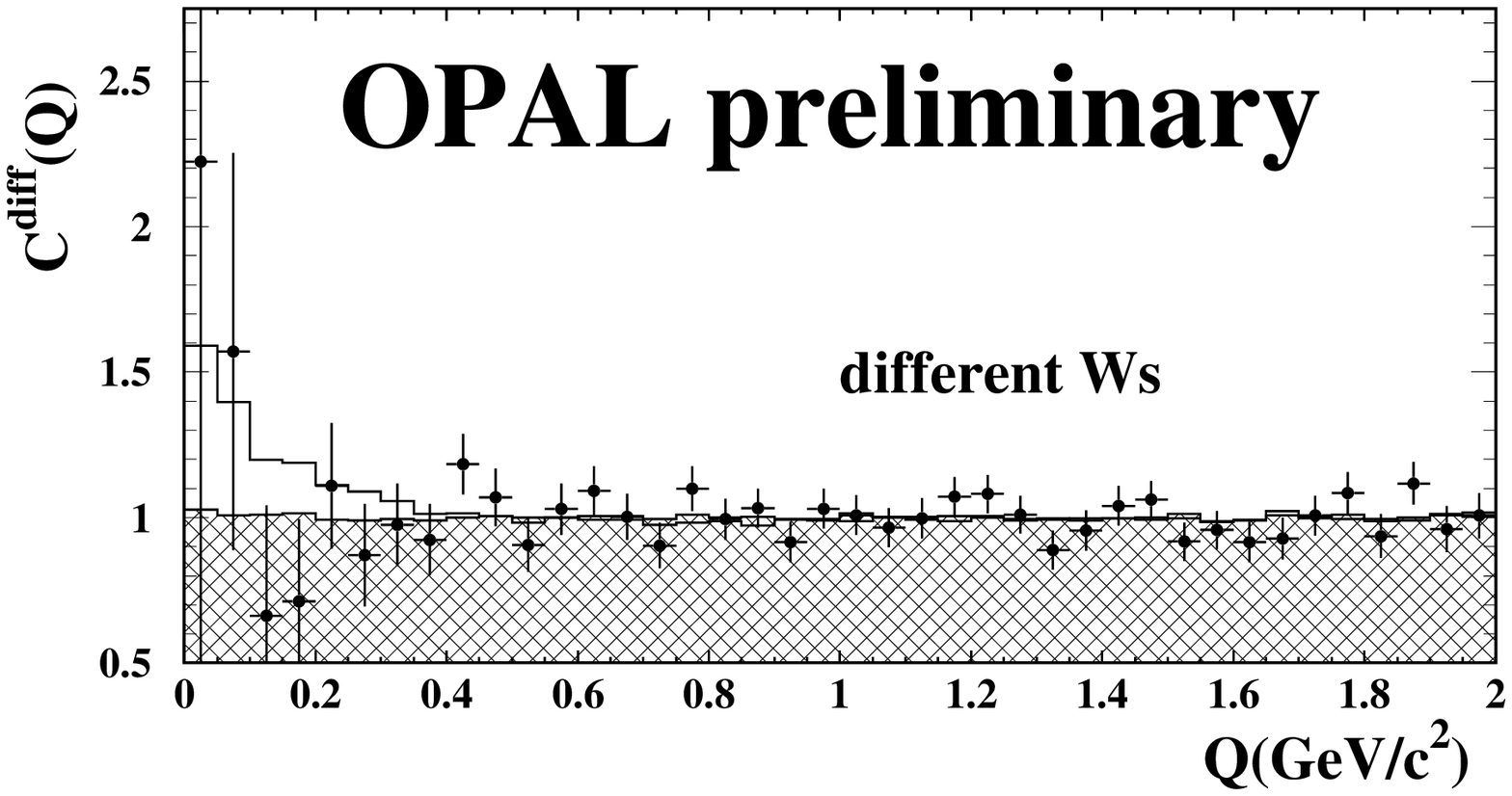,width=7.2cm}}
 \end{center}
 \vspace*{-3.4cm}
 \caption{Unfolded correlation for pions from different W's
          in OPAL\protect\cite{opal}.
          Data (crosses) are compared to MC
          with BEI-only (hatched),
          or BEI+BEB (open).
         }
 \vspace*{-7mm}
 \label{fig:opal}
\end{figure}

In OPAL\cite{opal},
$\mathrm{WW\rightarrow\ell\nu{q}\bar{q}}$, 
$\mathrm{WW\rightarrow{q}\bar{q}q\bar{q}}$ and 
non-radiative high-energy $\mathrm{(Z/\gamma)^*\rightarrow{q}\bar{q}}$ 
decays are analysed.
Correlation functions for 
these samples,
built from double ratios as in Eq.~(\ref{eq:double})
and adjusted for resonance multiplicities observed in data,
are unfolded as sums of contributions
from three categories:
pion pairs coming from the same W, from different W's,
or from hadronic Z decays.
BEC parameters for each category are determined in 
a simultaneous fit,
correcting for the differences in BEC between $\mathrm{q\bar{q}}$ events 
selected as (Z/$\gamma)^*$ and those selected as WW.
The BEC strength for the second category,
$\lambda^\mathrm{diff}$,
directly measures BEB 
and is expected to be zero in their absence.
MC events with BEC are only used 
for qualitative comparison to the data (Fig.~\ref{fig:opal}).
Fits are performed with different assumptions 
about BEC source sizes in the three categories.
The main fit yields
\be
\vspace*{-0.9mm}
\lambda^\mathrm{diff} = 0.05\pm0.67\pm0.35 \quad,
\vspace*{-0.4mm}
\ee
where systematics mainly come from tracking 
for low-$Q$ like-sign pairs
and the HERWIG simulation of resonances.
While the result is compatible with the absence of BEB, OPAL 
conclude that, at the current 
precision, they are unable to determine whether BEB exist.

\vspace*{-0.2mm}
\section{Conclusions}
\vspace*{-1mm}
Bose-Einstein correlations 
between like-sign charged pion pairs produced
in W-pair events at LEP
are studied by the four experiments 
using different techniques.
The existence of BEC
for pions coming from the same W is firmly established,
while no agreement is reached in the results of the four experiments
about the possible existence of BEC
between pions coming from different W's.
The existence of inter-W BEC
is supported by DELPHI at the 2$\sigma$ level,
whereas models of inter-W BEC are disfavoured
by L3 and ALEPH at the 4$\sigma$ and 2.2$\sigma$ level,
respectively.
No conclusion is reached by OPAL
at the present level of statistical accuracy.


\begin{thebibliography}{99}
\vspace*{-1mm}

\bibitem{gold}
  G.Goldhaber et al., {\it Phys. Rev.} {\bf 120} (1960) 300.

\bibitem{lepuno}
  ALEPH, {\it Phys.\hspace*{0.5mm}Rep.} {\bf 294} (1998) 1;
  DELPHI, {\it Phys. Lett.} B {\bf 471} (2000) 460;
  L3, {\it Phys. Lett.} B {\bf 458} (1999) 517;
  OPAL, CERN-EP/2000-004;
  and ref. therein. 

\bibitem{aleph}
  ALEPH 2000-039 CONF 2000-059, and ref.[1,2] therein.

\bibitem{delphi}
  DELPHI 2000-115 CONF 414, and ref.[16] therein.

\bibitem{l3}
  L3, CERN-EP/2000-084.

\bibitem{opal}
  OPAL PN393 (1999), and ref.[9] therein.

\bibitem{lonnblad} 
  L.L\"onnblad, T.Sj\"ostrand, {\it Eur. Phys. J.}
  C {\bf 2} (1998) 165, and ref. therein.

\bibitem{fourwmasses}
  ALEPH 2000-018 CONF 2000-015;
  DELPHI 2000-149 CONF 446;
  L3 Note 2575;
  OPAL PN422;
  and ref. therein.

\bibitem{lepwmass}
  The LEP WW Working Group, Note
  LEPWWG/MW/00-01 in preparation. 
  {\it http://lepewwg.web.cern.ch/LEPEWWG}.

\bibitem{bowler}
  M.G.Bowler, {\it Z. Phys.} C {\bf 29} (1985) 617,
  and ref. therein.

\bibitem{gyulassi}
  M.Gyulassi et al., 
  {\it Phys. Rev.} C {\bf 20} (1979) 2267.

\bibitem {kittel} 
  S.V.Chekanov, E.A. De Wolf and W. Kittel, 
  {\it Eur. Phys. J.} C {\bf 6} (1999) 403.

\bibitem{puk}
  N.Pukhaeva, these proceedings.

\end{thebibliography}
\end{document}